\documentclass[a4paper,11pt]{article}

\usepackage{jinstpub}

\usepackage[utf8]{inputenc}
\usepackage[T1]{fontenc}
\usepackage{lmodern}
\usepackage{microtype}

\usepackage{bm}
\usepackage{nicefrac}
\usepackage{siunitx}
\usepackage{graphicx}
\usepackage{caption}
\usepackage{subcaption}
\graphicspath{{figures/}{./}{../figures/}}
\usepackage{booktabs}
\usepackage{makecell}

\usepackage[inline]{enumitem}
\usepackage{paralist}
\usepackage{xspace}

\usepackage{url}
\usepackage{hyperref}

\usepackage{lineno}

\title{Test beam results for machine-learning-based reconstruction of resistive silicon sensors}

\author[1]{Don C. Wong,}
\author[2]{Leena Diehl,}
\author[1]{Spencer Ellis,}
\author[3]{Marcos Fern\'andez Garc\'ia}
\author[4]{Gabriele Giacomini,}
\author[2]{Vagelis Gkougkousis,}
\author[2]{Ben Kilminster,}
\author[1]{Jingyu Luo,}
\author[2]{Anna Macchiolo,}
\author[1]{Yajna Mahabeer,}
\author[2]{Yevhenii Padniuk,}
\author[2]{Daria Senina,}
\author[1]{Samantha Sunnarborg,}
\author[1]{Jessica Tang,}
\author[4]{Alessandro Tricoli,}
\author[1]{Lixing Wang,}
\author[1]{Gaetano Barone\thanks{Corresponding author.}}
       
\affiliation[1]{Brown University, 182 Hope Street, Providence, RI 02912, USA}
\affiliation[2]{University of Zurich}
\affiliation[3]{Instituto de F\'isica de Cantabria (CSIC-UC)}
\affiliation[4]{Brookhaven National Laboratory}

\date{\today}

\abstract{The advent of Low-Gain Avalanche Diodes (LGADs) and AC-coupled Low-Gain Avalanche Diodes (AC-LGADs) as candidate technologies for four-dimensional tracking has moved part of the reconstruction burden onto the analog waveforms of the sensor, in which the charge shared between electrodes encodes the impact position well below the electrode pitch. Recovering that position is a continuous, non-linear inverse problem. Transient Current Technique measurements characterize this problem at fixed injected charge and high signal-to-noise, whereas a minimum-ionizing beam adds Landau fluctuations in the deposited charge under detector operating conditions. We report test beam results for full-waveform reconstruction of an AC-LGAD of \SI{500}{\micro m} pitch read out on a $2\times2$ pad matrix, exposed to the \SI{120}{GeV} pion beam of the CERN Super Proton Synchrotron (SPS) and read out alongside a six-plane MIMOSA-26 telescope. The telescope is aligned, and its per-plane resolutions are calibrated in situ, giving a track pointing resolution of about \SI{2.5}{\micro m} per axis at the device, so that the measured residual is the sensor response rather than a convolution with the reference. We then compare two analytic estimators, the parameter-free charge-imbalance formula and a fitted linear matrix inversion, with the topology-agnostic transformer developed for resistive silicon sensors. Over the full active pad, the transformer reaches a median total position with up to factor 3 improvement. Within the charge-sharing core, the three methods converge to the floor of this sensor at the achieved signal-to-noise. }

\begin{document}
\maketitle
\flushbottom

%

\section{Introduction}
\label{sec:intro}

Precision timing combined with precision position measurement is a requirement of the next generation of collider experiments, where four-dimensional tracking suppresses pile-up, enables time-of-flight particle identification, and resolves dense vertices~\cite{cms_tdr_mtd,cmscollaboration_2008,ecfa_roadmap_2021}. Similar requirements arise in medical applications~\cite{royon2024168886} and in space instrumentation~\cite{gautier2021165599,wu20192672}. LGADs deliver time resolutions of a few tens of picoseconds through a thin, moderately doped multiplication layer~\cite{pellegrini2014,wu_2020,cartiglia2020,Sadrozinski2018,lgad1,lgad2}, but their segmentation relies on a continuous gain layer interrupted by inter-pad regions that are dead to signal, so spatial information is quantized at the pad pitch. In the AC-coupled variant, also called the Resistive Silicon Detector (RSD), the AC pad is coupled to a resistive $n^+$ layer through a dielectric and the gain layer is left unsegmented~\cite{mandurrino2020_nima959,tornago2021_nima1003_165319}. The result is a 100\% fill factor and a signal that is shared among several neighboring electrodes according to the impact position, so that a single device is simultaneously a precision-timing and a precision-position detector~\cite{apresyan2020,heller2022_aclgad_bnl_hpk_jinst,damen222}.

The same sharing that makes interpolation possible complicates the readout. Position reconstruction becomes a continuous inverse problem, mapping a vector of correlated, noisy analog waveforms onto an impact coordinate. The mapping is neither exactly linear, since sharing saturates as the impact point approaches an electrode, nor stationary, since it depends on bias voltage, gain and local electrode topology. Inverting the pad response matrix directly is the information-optimal approach but becomes ill-conditioned once off-diagonal elements are large, and centroid or charge-imbalance estimators regularize those fluctuations at the cost of discarding information. Machine learning has therefore been introduced to regularize the reconstruction while retaining the full waveform. Recent results show dense networks reaching position resolutions of \SIrange{2}{10}{\micro m} at \SIrange{100}{200}{\micro m} pitch~\cite{siviero2021}, random forests reaching $\approx\SI{8}{\micro m}$ at \SI{200}{\micro m} pitch~\cite{tornago2023}, neural regression in beam tests delivering \SIrange{5}{10}{\micro m} on strips~\cite{heller2022_aclgad_bnl_hpk_jinst,madrid2023}, $\sigma_x\approx\SI{15}{\micro m}$ on a \SI{450}{\micro m} pitch matrix read out by a dedicated ASIC~\cite{menzio2024}, and a laboratory-trained network validated in a beam reaching \SI{65}{\micro m} on a \SI{450}{\micro m} pitch RSD, a 50\% gain over an analytic model~\cite{investigationchargesharing}. In a companion study we introduced recurrent and topology-agnostic transformer architectures for full-waveform reconstruction and demonstrated approximately \SI{10}{\micro m} resolution on $500\times\SI{500}{\micro m\squared}$ pitch sensors using transient-current technique (TCT) laser data~\cite{aoki2026}, building on earlier work on machine-learning-processed readout and waveform compression for signal-sharing devices~\cite{baronedrd3,barone2025}.

Laser data do not reproduce the detector operating regime. A TCT scan injects a fixed, calibrated charge at a known point on a fine grid, so both the label noise and the charge fluctuation are small by construction. A minimum-ionizing particle deposits a Landau-distributed charge at an unknown position, and the position label must be supplied by an independent reference detector. This work closes that gap. We expose an AC-LGAD of \SI{500}{\micro m} pitch, read out on a $2\times2$ pad matrix, to the \SI{120}{GeV} pion beam of the CERN SPS, read it out as full waveforms alongside a six-plane MIMOSA-26 telescope, and evaluate the transformer of Ref.~\cite{aoki2026} against analytic baselines on the same matched events. Preliminary results from this dataset, using a recurrent architecture, were reported in Ref.~\cite{wong2025rnn}. Section~\ref{sec:setup} describes the setup, Section~\ref{sec:tracking} the track reconstruction and the telescope performance that sets the accuracy of the position labels, Section~\ref{sec:sharing} the charge sharing measured in the device, and Section~\ref{sec:reco} the reconstruction methods and their results.

\section{Sensor and experimental setup}
\label{sec:setup}

Data were recorded in October 2025 in the North Area of the CERN SPS with a secondary hadron beam of \SI{120}{GeV}$/c$ pions. At this momentum the multiple-Coulomb-scattering angle in the telescope material scales as $1/p$ and is roughly a factor twenty smaller than for the few-GeV electron beams of lower-energy facilities. The scattering contribution to the track fit is therefore sub-dominant, and the telescope pointing is limited by the intrinsic pixel resolution.

\begin{figure}[tbp]
  \centering
  \includegraphics[width=\textwidth]{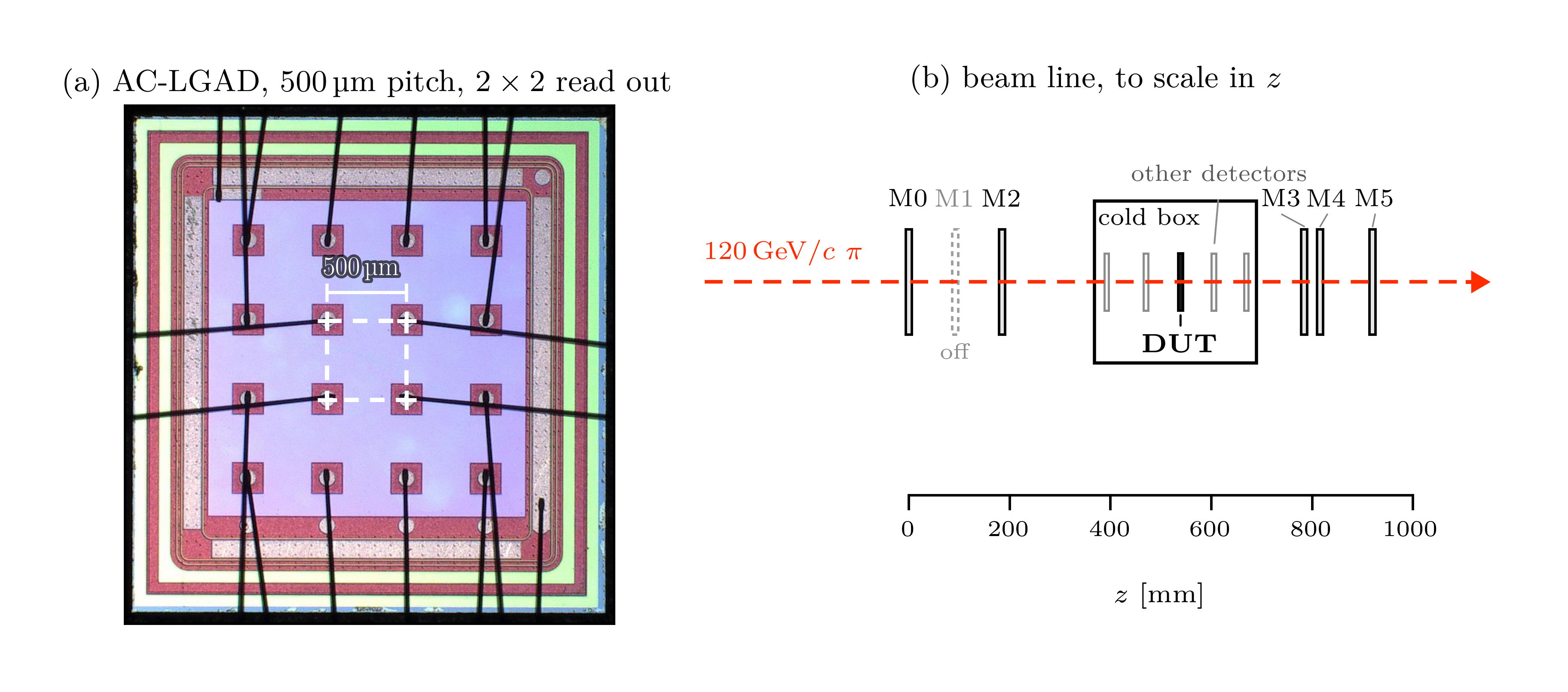}
  \caption{(a) Optical micrograph of a BNL AC-LGAD of the type used here. The array is $4\times4$ pads on a \SI{500}{\micro m} pitch. The dashed white square is exactly one $500\times\SI{500}{\micro m\squared}$ readout cell. The wire bonds of the grounded pads are also visible. (b) The beam line inside the cold box. The five active MIMOSA-26 planes are at $z=0$, $185$, $784$, $816$ and \SI{920}{mm}, the disabled plane M1 is drawn in gray between M0 and M2, and the DUT is at $z=\SI{539}{mm}$. Four other detectors were read out in the same runs but are not used in this analysis.}
  \label{fig:setup}
\end{figure}

The device under test (DUT) is an AC-LGAD manufactured at Brookhaven National Lab (BNL) from the same production and characterization campaign as Ref.~\cite{aoki2026}. It carries a $4\times4$ array of \SI{200}{\micro m} pads on a \SI{500}{\micro m} pitch; the central $2\times2$ pads are wire bonded to the readout and the remaining pads are connected to ground, which is the configuration shown in Fig.~\ref{fig:setup}(a). Assigning a hit to the pad with the largest signal, without using any sharing information, would give a resolution of pitch$/\sqrt{12} \approx \SI{144}{\micro m}$ per axis, and this is the figure that charge sharing has to beat. Table~\ref{tab:sensor} reproduces the fabrication parameters reported in Ref.~\cite{aoki2026} together with the two entries specific to this device. Several other detectors of different topologies were mounted in the same beam, and the reconstruction methods described below are designed to be agnostic to pad topology.

\begin{table}[tbp]
  \centering
  \caption{BNL AC-LGAD sensor information.}
  \label{tab:sensor}
  \begin{tabular}{ll}
    \toprule
    Item & Information\\
    \midrule
    Producer & Brookhaven National Laboratory\\
    Implant type & Uniform p-type implants on the silicon surface\\
    Active area & \SI{1.3}{mm\squared}\\
    Gain-layer dose, \SI{30}{\micro m} sensors & $2.7\times10^{12}\,\mathrm{cm}^{-2}$\\
    Gain-layer dose, \SI{20}{\micro m} sensors & $2.25\times10^{12}\,\mathrm{cm}^{-2}$\\
    Baseline geometry & Square grid\\
    Secondary geometries & Diamond, triangular, and other geometries\\
    \midrule
    Pad array and pitch of the DUT & $4\times4$ pads, \SI{200}{\micro m} pad, \SI{500}{\micro m} pitch\\
    Instrumented pads & central $2\times2$, remaining pads grounded\\
    \bottomrule
  \end{tabular}
\end{table}

The trajectory of each beam particle is measured by a telescope of six MIMOSA-26 monolithic active pixel sensors~\cite{huguo2010,jansen2016}, each a $1152\times576$ matrix of \SI{18.4}{\micro m} pixels with binary readout and a nominal single-hit resolution near pitch$/\sqrt{12} \approx \SI{5.3}{\micro m}$. Two planes sit upstream of the device region, at $z=0$ and \SI{185}{mm}, and three downstream, at $z=784$, $816$ and \SI{920}{mm}; the sixth plane is disabled because of a persistent noise pathology, which leaves five planes in the fit. The MIMOSA-26 has a \SI{230}{\micro s} rolling-shutter integration time, much longer than the beam inter-particle spacing, so a single triggered frame typically contains several superimposed particles and the track finder has to resolve real tracks against a combinatorial background. The DUT is mounted at $z=\SI{539}{mm}$, between the two telescope arms, where the pointing resolution is best. The complete beam line is drawn to scale in Fig.~\ref{fig:setup}(b).

The four instrumented AC-LGAD pads are digitized as full waveforms by a CAEN DT5748 sampling digitizer at 5\,GS/s on channels 12 to 15. Waveforms are baseline-subtracted on their first 100 samples and a fixed window of 130 samples around the pulse is retained. The telescope planes and the digitizer are read out through EUDAQ2~\cite{ahlburg2020} and synchronized event by event through a common Trigger Logic Unit, with the trigger identifiers of the two sub-systems cross-checked for every event.

Five runs were analyzed, in two bias-voltage groups (Table~\ref{tab:runs}). Gain grows with bias, and the median leading-pad amplitude rises from \SIrange{14.9}{15.3}{mV} at \SI{135}{V} to \SI{19.9}{mV} at \SI{160}{V}, against a baseline noise of \SI{1.01}{mV} rms per channel. The two high-statistics \SI{160}{V} runs are the position-reconstruction dataset we use, and they give \num{59026} matched track-waveform pairs between them.

\begin{table}[tbp]
  \centering
  \caption{The five October runs. ``Matched pairs'' are tracks that fall inside the DUT active area and are accompanied by a pulse above the signal threshold.}
  \label{tab:runs}
  \begin{tabular}{lccccc}
    \toprule
    Run & Bias (V) & Events & Accepted tracks & med.\ $\chi^2/\mathrm{ndof}$ & Matched pairs\\
    \midrule
    012627 & 135 & 12\,565 & 12\,137 & 0.90 & 2\,355\\
    012631 & 135 & 13\,906 & 13\,498 & 0.91 & 2\,505\\
    012632 & 160 & 26\,654 & 25\,773 & 0.93 & 4\,071\\
    012640 & 160 & 142\,448 & 138\,728 & 1.07 & 33\,934\\
    012653 & 160 & 111\,557 & 108\,457 & 1.08 & 25\,092\\
    \bottomrule
  \end{tabular}
\end{table}

\section{Track reconstruction and telescope performance}
\label{sec:tracking}

The resolution of the sensor can only be measured to the extent that the reference telescope points at it, so we characterize the telescope before quoting any number for the DUT.

The EUDAQ2 binary is decoded into pixel hits, per-channel waveforms and trigger timestamps. Pixels whose hit rate exceeds a fixed multiple of the plane mean are masked, which keeps the seed combinatorics of the track finder tractable in the high-multiplicity rolling-shutter frames. Correlation residuals against the reference plane then remove the gross plane-to-plane offsets. The five active planes are aligned by minimizing the global track $\chi^2$, first in the translations and the in-plane rotation and then over all five degrees of freedom per plane. A weak Gaussian prior on the tilt angles stabilizes the poorly constrained out-of-plane rotations~\cite{blobel2006}. The reference plane is held fixed to define the frame, and the rotation convention is the extrinsic $R_zR_yR_x$ sequence throughout, following Ref.~\cite{dannheim2021}. With the telescope frozen, the pose of the DUT is obtained by matching telescope-extrapolated impact points to the charge centroids of the four pads. This proceeds through a median-shift pre-step, a closed-form Kabsch rigid-body fit and a Millepede-style linear refinement. An $\eta$-correction built from the observed charge-asymmetry distribution then linearizes the response near the electrodes, and the pose is refit. We refer to this chain as \textsc{T\textsuperscript{3}} (Tabulated Telescope Tracking, or triple-T): rather than refitting every track at each step of the alignment minimization, each track is reduced once per iteration to a small set of sufficient statistics, from which the global $\chi^2$ and its dependence on the pose of a plane are evaluated in closed form. The alignment therefore runs on the full event sample rather than on a subsample, already implicit in an iterative alignment. The full chain used in Section~\ref{sec:reco} is implemented in the \texttt{tb\_tracking} package~\cite{tbtracking2026}.

Tracks are reconstructed through the aligned telescope by seeding on hit pairs and completing plane by plane, and are fit by a weighted least-squares straight line in three dimensions using the per-plane, per-axis intrinsic resolution as the measurement weight, with a Highland multiple-scattering term appropriate to \SI{120}{GeV} pions~\cite{highland1975,kleinwort2012}. Each fitted track is extrapolated to the plane of the DUT and joined to the four waveforms of the same event, and these matched pairs are the inputs of Section~\ref{sec:reco}.

\begin{figure}[tbp]
  \centering
  \includegraphics[width=\textwidth]{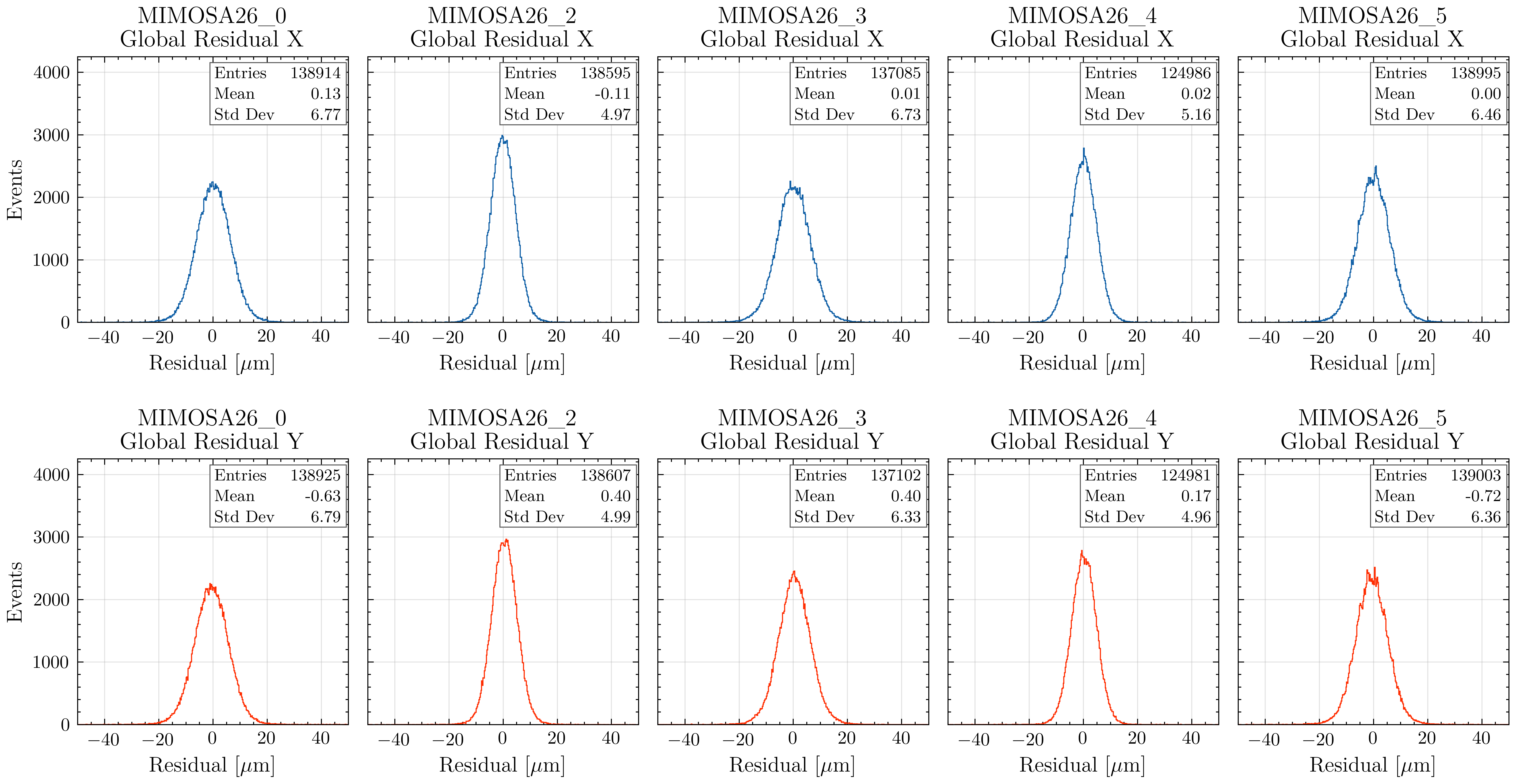}
  \caption{Telescope residuals per plane after alignment, in $x$ (top row) and $y$ (bottom row), for run 012640. The distributions are zero-mean and free of position-dependent tilt, and their widths give the per-plane intrinsic resolutions used as fit weights.}
  \label{fig:telres}
\end{figure}

\begin{figure}[tbp]
  \centering
  \includegraphics[width=\textwidth]{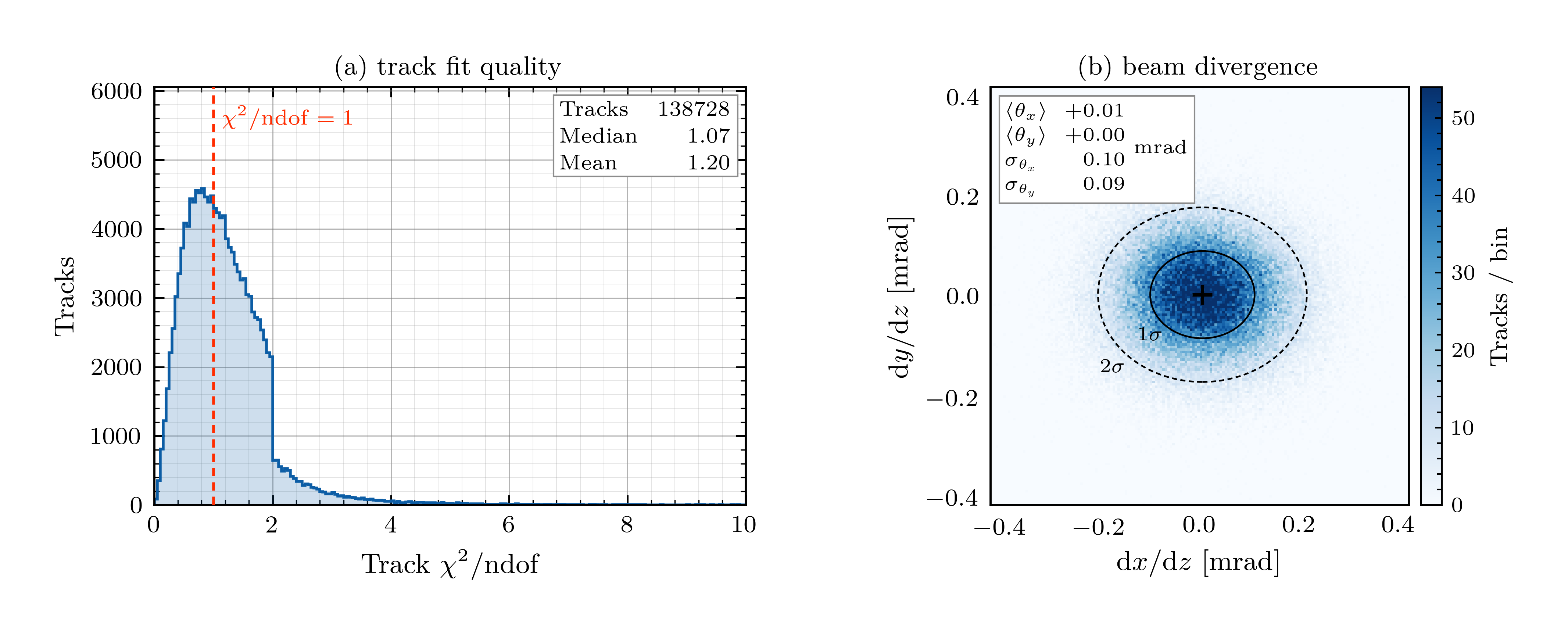}
  \caption{Quality of the reference tracks in run 012640. (a) Track $\chi^2/\mathrm{ndof}$, which peaks at unity once the calibrated per-plane resolutions are used as fit weights; tracks with $\chi^2/\mathrm{ndof}>10$ are rejected. (b) Joint distribution of the fitted track slopes, whose width is the angular spread of the pion beam through the telescope.}
  \label{fig:tracking}
\end{figure}

After alignment, \textsc{T\textsuperscript{3}} recovers the per-plane, per-axis intrinsic resolution iteratively from the leave-one-out residual widths, giving \SIrange{3.8}{5.4}{\micro m} across the five planes, consistent with the MIMOSA-26 binary-readout expectation of pitch$/\sqrt{12} \approx \SI{5.3}{\micro m}$. These calibrated resolutions are written back into the geometry and used as the measurement weights in the final fit. Iterating the calibration drives the track $\chi^2/\mathrm{ndof}$ to unity (Fig.~\ref{fig:tracking}(a) and Table~\ref{tab:runs}) and leaves residuals that are zero-mean and free of position-dependent tilt (Fig.~\ref{fig:telres}). The fitted track slopes give a beam divergence of \SI{0.10}{mrad} in $x$ and \SI{0.09}{mrad} in $y$ (Fig.~\ref{fig:tracking}(b)), narrow enough that the incidence angle contributes nothing to the residual at the thickness of this sensor.

\section{Charge sharing in the DUT}
\label{sec:sharing}

Before comparing reconstruction methods it is useful to look at the response the methods have to invert. We work in the local frame $(x,y)$ of the DUT, with the origin at the center of the readout cell, and combine the two \SI{160}{V} runs using each run's own pose. The analysis fiducial is $|x|,|y| < \SI{0.50}{mm}$, one full pitch either side of the origin. It is symmetric about the readout cell and extends half a pitch beyond the cell on every side, so the behavior outside the cell can be measured. It also stays inside the $1.1\times\SI{1.1}{mm\squared}$ active area of the sensor. \num{36712} of the \num{59026} matched pairs fall inside it.

\begin{figure}[tbp]
  \centering
  \begin{minipage}[b]{0.52\textwidth}
    \centering
    \includegraphics[width=\textwidth]{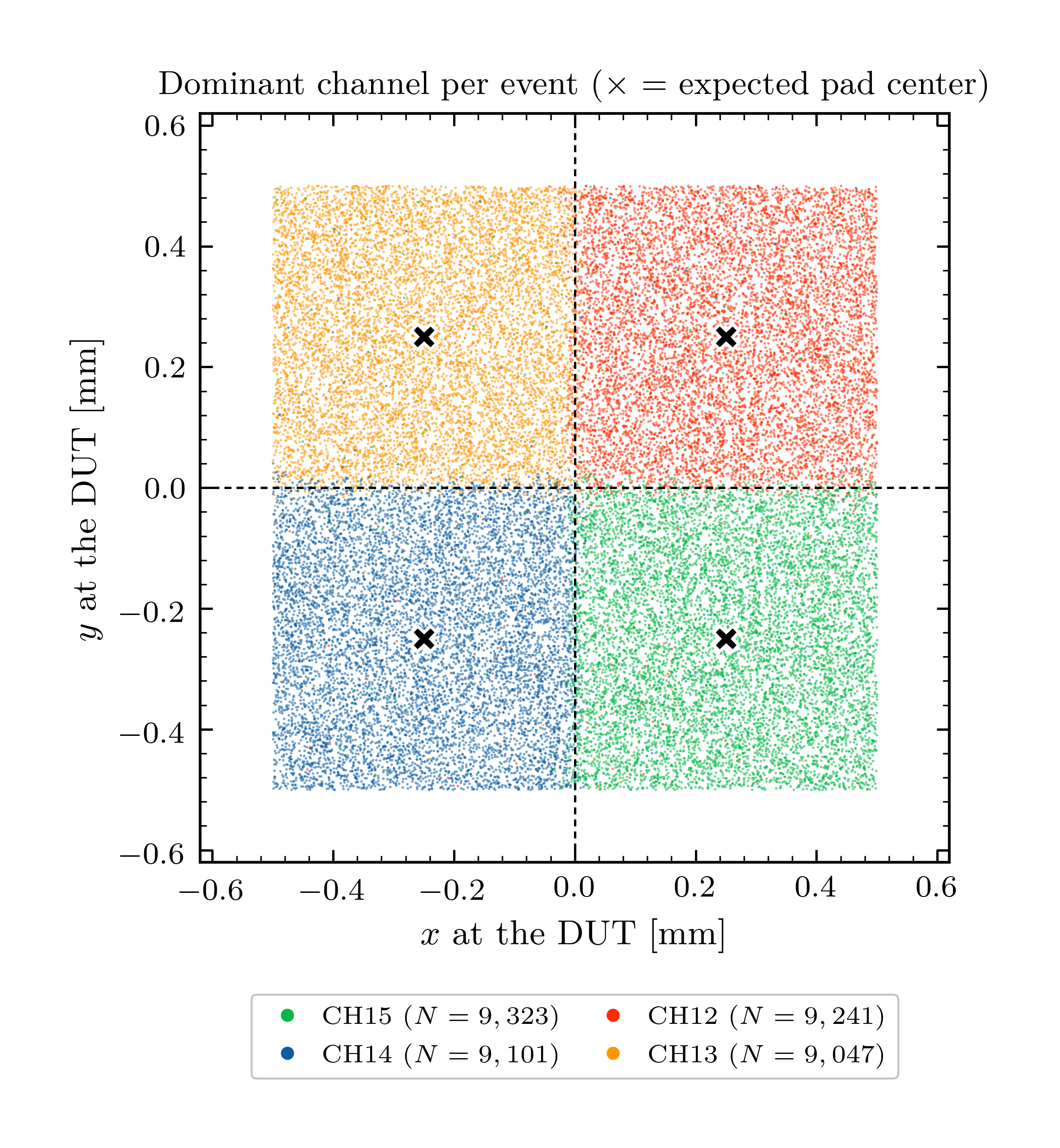}\\[2pt]
    {\small\textbf{(a)}}
  \end{minipage}\hfill
  \begin{minipage}[b]{0.46\textwidth}
    \centering
    \includegraphics[width=\textwidth]{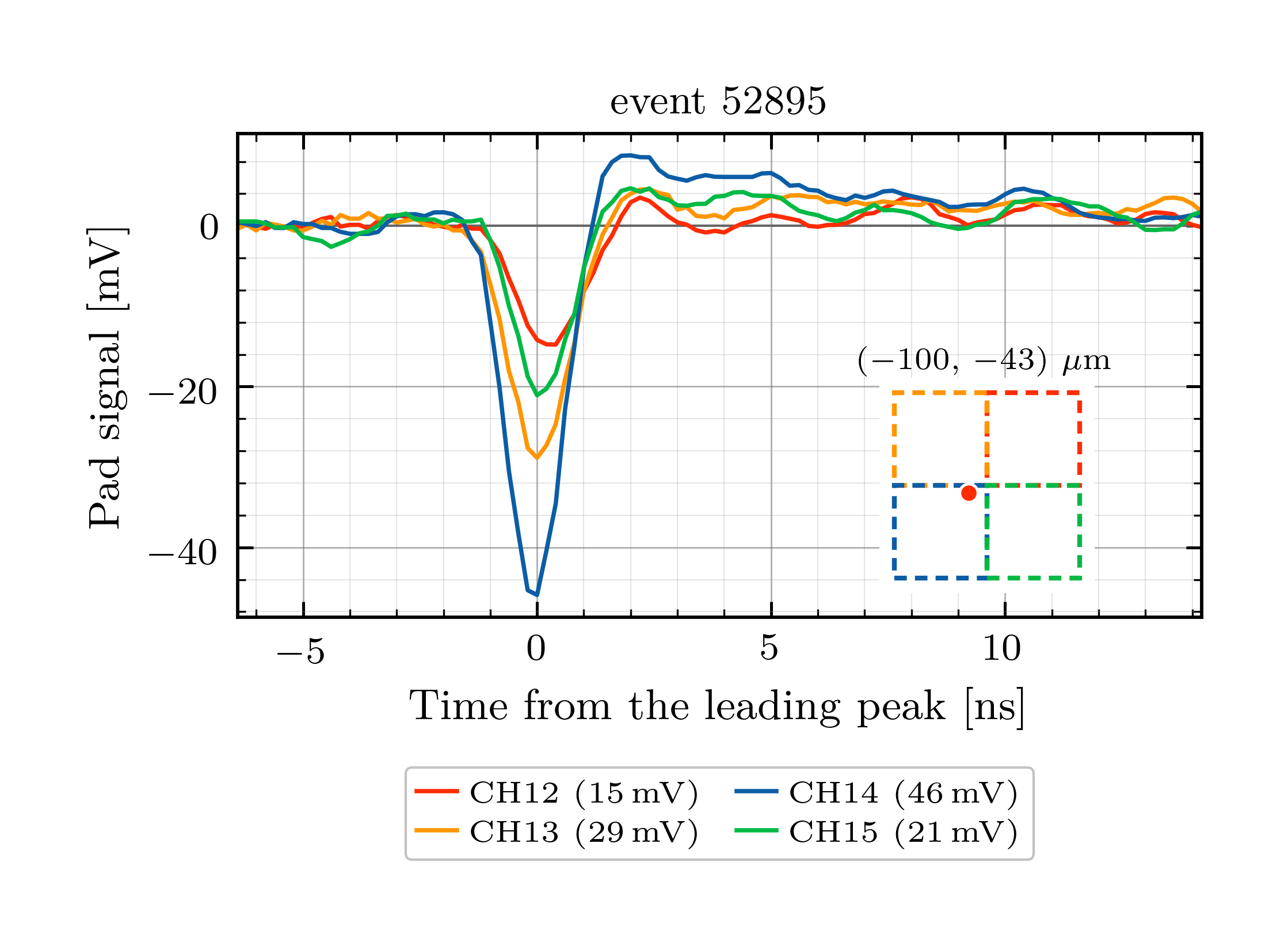}\\[2pt]
    {\small\textbf{(b)}}
  \end{minipage}
  \caption{(a) Channel with the largest pulse amplitude, for each of the \num{36712} matched events of the combined \SI{160}{V} dataset inside the analysis fiducial, drawn at the impact point measured by the telescope. Crosses mark the pad centers implied by a \SI{500}{\micro m} pitch, which coincide with the centers of the four domains. The domains meet on $x=0$ and $y=0$ and contain equal numbers of events to about 1.5\%. (b) The four baseline-subtracted pad waveforms of a single event. The inset shows where the telescope says the particle hit, relative to the four readout pads drawn in the same colors as the waveforms.}
  \label{fig:device}
\end{figure}

Fig.~\ref{fig:device}(a) colors every one of those events by the channel with the largest pulse amplitude. The four sections are obvious, they meet on the lines $x=0$ and $y=0$, and each is centered on the pad center implied by a \SI{500}{\micro m} pitch. The map uses no reconstruction and no fit. It is therefore a direct check that the pose of the DUT and the pad-to-channel assignment are both correct, and that the four channels have comparable gain. The four domains contain between \num{9047} and \num{9323} events, a spread of about 1.5\%.

The pitch can be measured from the same data. Writing $\eta_x = (Q_R - Q_L)/(Q_R + Q_L)$ for the charge asymmetry between the right and left pad sums, a least-squares fit of $x = k\,\eta_x$ restricted to $|x|,|y| < \SI{0.15}{mm}$, where the response is close to linear, gives $k_x = \SI{245.6}{\micro m}$ and $k_y = \SI{249.7}{\micro m}$. For a symmetric $2\times2$ array the scale $k$ is half the pad pitch, so the data imply a pitch of \SI{495}{\micro m}, in agreement with the layout. We use the nominal \SI{500}{\micro m} throughout, and in particular we fix $k = \mathrm{pitch}/2 = \SI{250}{\micro m}$ in the analytic baseline below rather than treating it as a free scale.

\section{Position reconstruction from AC-LGAD waveforms}
\label{sec:reco}

The reconstruction problem is to map the four pad waveforms of an event onto the impact coordinate $(x,y)$. We score every method on the identical held-out events so that the comparison is like for like, and we quote the total, two-dimensional residual $\Delta = \sqrt{(\hat x - x)^2 + (\hat y - y)^2}$, summarized by its median and by its $90^{\rm th}$ and $99^{\rm th}$ percentiles. This is the figure of merit used in Ref.~\cite{aoki2026}. For comparison with prior work, we also report the per-axis median absolute residual and the $16$--$84\%$ half-width. The network was trained on the wider $|x|,|y| < \SI{0.585}{mm}$ sample of \num{49732} matched pairs, split 60/20/20 into \num{29839} for training, \num{9946} for validation and \num{9947} held out. Every number quoted below is evaluated on the $N_{\rm test} = \num{7392}$ of those held-out events that fall inside the \SI{0.50}{mm} fiducial. No event seen in training therefore enters any figure of merit, and the network is never asked to extrapolate beyond the region it was trained on.

\subsection{Analytic baselines}

The simplest baseline is a binary readout, which assigns the hit to the center of the pad with the largest amplitude and uses no sharing information at all. It is the reference that any interpolation has to beat.

The parameter-free interpolating estimator is the charge-imbalance, or $\eta$, formula. With $Q_i$ the peak amplitude on pad $i$ and $L,R,T,B$ the left, right, top and bottom pad sums,
\begin{equation}
  \hat{x} = k\,\frac{Q_R-Q_L}{Q_R+Q_L}, \qquad
  \hat{y} = k\,\frac{Q_T-Q_B}{Q_T+Q_B},
  \label{eq:eta}
\end{equation}
with $k = \mathrm{pitch}/2 = \SI{250}{\micro m}$ fixed by the geometry as established in Section~\ref{sec:sharing}. For a symmetric $2\times2$ array this is algebraically identical to the amplitude-weighted centroid over pad coordinates $\pm k$. It requires no fit and no training data. It is important that $k$ be pinned to the geometry rather than tuned per selection, since an estimator whose scale is refitted for every region is no longer a baseline. Note that Eq.~(\ref{eq:eta}) is bounded by construction, $|\hat x| \le k$, so it cannot reconstruct a hit that falls outside the readout cell.

The second estimator inverts the linear pad response. Writing $f_i = Q_i/\sum_j Q_j$ for the charge fractions,
\begin{equation}
  \begin{bmatrix} \hat x & \hat y \end{bmatrix}
  = \begin{bmatrix} f_1 & f_2 & f_3 & f_4 & 1 \end{bmatrix} B,
  \label{eq:matrixinv}
\end{equation}
with the ten coefficients of $B$ obtained by a Moore-Penrose least-squares solve on the \num{29320} events inside the fiducial that are not held out, and applied to the test split. The two interpolating estimators differ in exactly one respect, whether the charge-to-position scale is taken from geometry or solved for from data. Both share the same limitation. Binning the events in $x$ shows that the charge asymmetry $\eta_x = (Q_R-Q_L)/(Q_R+Q_L)$ rises monotonically with $x$ only inside the readout cell. It peaks at $|\eta_x| \approx 0.66$ near $|x| = \SI{0.30}{mm}$, where the two far pads have fallen to their \SI{3}{mV} floor, and then turns over. Beyond that point the asymmetry is two-valued, so any monotonic function of it has to fold its estimates back towards the pad centers.

\subsection{Learned estimator}

The learned estimator is the topology-agnostic transformer developed for resistive silicon sensors in Ref.~\cite{aoki2026}, applied here without architectural change to test beam data. A linear embedding projects each channel's waveform and its two pad coordinates into a shared latent space. The resulting tokens are processed by a stack of six transformer blocks~\cite{vaswani2017} with multi-head self-attention, layer normalization and residual connections. Each pad therefore attends to every other pad, and the cross-channel correlations, which are part of the physical response, are available to the network. Because attention supports an arbitrary number of tokens, the same network accepts any pad multiplicity or topology. The adaptive layer normalization is conditioned on the bias voltage and is designed to allow one model to span operating points. A position head maps the final embedding to a scalar weight per pad and returns a weighted sum of the pad coordinates. The weights come from a plain linear layer and are not constrained to a simplex, so the prediction can leave the convex hull of the pad centers and reach the edges of the fiducial. We use the variant that predicts a per-event covariance and is trained by the corresponding negative log-likelihood, which yields a calibrated error estimate and stabilizes training. The model has \num{406336} parameters and is trained for at most 400 epochs with early stopping on the validation error. Waveforms are baseline-subtracted, gain-equalized across channels, cropped to a 130-sample window around the peak and normalized using training-split statistics only.

\subsection{Results}

\begin{figure}[tbp]
  \centering
  \includegraphics[width=\textwidth]{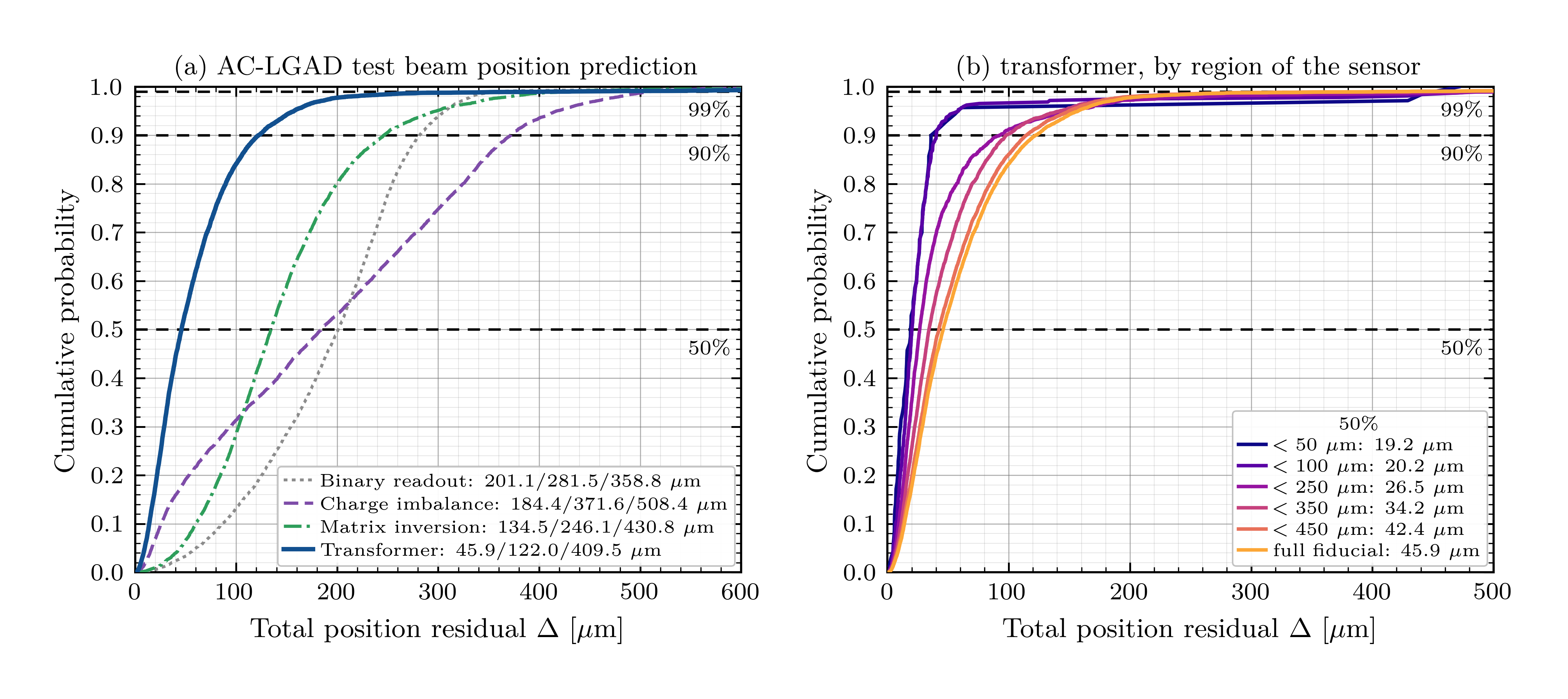}
  \caption{Cumulative probability of reconstructing a hit as a function of the total position residual. (a) All estimators on the identical held-out events over the full fiducial. The legend gives the median, $90^{\rm th}$ and $99^{\rm th}$ percentiles of each, and the dashed black lines mark those three levels. (b) The deployed transformer alone, evaluated on nested square regions of the sensor, so that the price of leaving the \SI{500}{\micro m} readout cell can be read off directly.}
  \label{fig:cdf}
\end{figure}

Over the full fiducial the transformer reaches a median total residual of \SI{45.9}{\micro m}, against \SI{134.5}{\micro m} for matrix inversion and \SI{201.1}{\micro m} for the binary readout, a factor $2.9$ improvement over the better analytic estimator and a factor $4.4$ over the binary readout. Expressed per axis, the median absolute residual is \SI{24.2}{\micro m}, or $4.8\%$ of the pitch, and the $16$--$84\%$ half-width of the per-axis residual is \SI{41.1}{\micro m}. Fig.~\ref{fig:cdf}(a) shows the cumulative distributions of every estimator over the full fiducial, and Fig.~\ref{fig:cdf}(b) shows the transformer alone on nested regions of the sensor.

Charge imbalance is the worst interpolating estimator over the full fiducial, at \SI{184.4}{\micro m}, barely better than the binary readout. This is not a defect of the formula but a statement about the region. Eq.~(\ref{eq:eta}) is bounded by $k = \SI{250}{\micro m}$, and 74\% of the test events fall outside that range, so every hit beyond the readout cell is pulled back towards the cell boundary. The same effect appears in the linear estimator in a milder form: inside the readout cell the matrix inversion tracks the true coordinate, and outside it the reconstructed coordinate turns over and folds back towards the pad centers, reproducing the turnover of the charge asymmetry itself.

The contribution of waveform information can be estimated by comparison with an amplitude-only model. A gradient-boosted regression trained on the four peak amplitudes and their sum, on the same split, gives \SI{76.9}{\micro m} over the full fiducial. Dropping the assumption of linearity while staying in the amplitude domain therefore gains a factor $1.8$ over Eq.~(\ref{eq:matrixinv}), and the full-waveform transformer gains a further factor $1.7$. This additional gain is consistent with useful information in the pulse shape, although the comparison also changes the model class. Beyond the turnover the asymmetry is two-valued, so an amplitude-only estimator has to break the degeneracy using the falling total amplitude, which is the weaker of the two handles.

Inside the readout cell the picture changes. The transformer improves to \SI{26.5}{\micro m} and the parameter-free $\eta$ formula to \SI{32.4}{\micro m}, so the gap narrows from a factor $2.9$ to a factor $1.2$ as the region shrinks and the charge-sharing response becomes close to linear. The transformer's advantage is therefore entirely a large-region effect. It is level with the analytic estimators inside $\max(|x|,|y|) < \SI{0.15}{mm}$ and pulls away only once the response starts to saturate.

\subsection{Limits to the resolution}

The telescope contributes about \SI{2.5}{\micro m} per axis and cannot account for the core figure in quadrature. To test whether the pulse height does, we split the \num{713} core events into eight equal-population bins of the event signal-to-noise ratio $\mathrm{SNR} = \sum_i A_i / \sqrt{\sum_i \sigma_i^2}$. Here $A_i$ is the peak amplitude and $\sigma_i$ the baseline noise of channel $i$. In each bin we measure the $16$--$84\%$ half-width of the per-axis residual of the parameter-free estimator, pooling $x$ and $y$. A least-squares fit of those eight points, linear in $1/\mathrm{SNR}^2$, gives
\begin{equation}
  \sigma^2 = \left(\frac{242}{\mathrm{SNR}}\right)^2 + \left(\SI{11.5}{\micro m}\right)^2 ,
  \label{eq:snr}
\end{equation}
and the median event has $\mathrm{SNR} = 19.1$ over the whole fiducial and $\mathrm{SNR} = 20.3$ in the core, which places the measurement on the steep part of the curve. Doubling the signal-to-noise ratio would take the per-axis width from the \SI{16.6}{\micro m} that Eq.~(\ref{eq:snr}) gives at the median of this sample to \SI{12.9}{\micro m}, and a further doubling would gain less than \SI{1}{\micro m}. This is the essential difference from the laser regime of Ref.~\cite{aoki2026}, where the injected charge is fixed and calibrated, and it accounts for the gap between the two measurements in the core. The asymptote of Eq.~(\ref{eq:snr}), \SI{11.5}{\micro m} per axis, is consistent with the approximately \SI{10}{\micro m} reached on the same pitch with a laser. That asymptote is a fitted intercept rather than a measurement: the highest bin reaches only $\mathrm{SNR} \approx 33$, so it is extrapolated beyond the data, and it moves by \SI{0.6}{\micro m} if the upper 3\% of the signal-to-noise distribution is trimmed. We also did not vary the deposited charge, so we cannot separate the Landau fluctuation from the electronic noise here; what the data show is that the two regimes agree once the pulse height is conditioned on.

A separate control isolates the role of the training region from that of the evaluation region. Four models were trained on progressively larger regions of the sensor and scored on one fixed test set spanning the wider \SI{0.585}{mm} sample, so the numbers in this paragraph are on that wider sample rather than on the \SI{0.50}{mm} fiducial used elsewhere. The core resolution barely depends on where the model was trained, at \SIrange{17}{20}{\micro m} whether the training region was the core, the middle or the whole fiducial, which confirms that the core figure belongs to the region and not to the training selection. Restricting the training to the core, on the other hand, substantially degrades performance outside it. Such a model gives \SI{307}{\micro m} over the full fiducial and \SI{463}{\micro m} in the outermost shell, against \SI{54.5}{\micro m} and \SI{71.3}{\micro m} for the model trained on the full region. The network does not extrapolate past its training region, so the deployable configuration is the model trained on the full region, and the two figures of merit should be reported and labeled separately. Within the core the run-to-run spread of eight trainings that differ only in the weight-initialization seed is $\pm\SI{0.5}{\micro m}$. A scan of the pad-coordinate scale over a factor $4.4$ moves the median by less than \SI{1.1}{\micro m}, which is inside that band, so we do not quote a tuned optimum.

\subsection{Comparison with other methods}

Normalized to the pitch, the result is competitive with published beam-test reconstruction of resistive silicon sensors. Inside one readout cell, which is the region such measurements normally cover, the transformer gives a per-axis residual half-width of \SI{24.0}{\micro m}, or $4.8\%$ of the \SI{500}{\micro m} pitch, falling to \SI{17.6}{\micro m}, or $3.5\%$, in the core. A \SI{450}{\micro m} pitch matrix read out by a dedicated ASIC reaches $\sigma_x \approx 3.3\%$ of pitch~\cite{menzio2024}, and thin resistive silicon detectors with large pixels reach comparable fractions~\cite{arcidiacono2023}. A laboratory-trained network on a \SI{450}{\micro m} pitch RSD validated at DESY reaches \SI{65}{\micro m}, or $14.4\%$ of pitch, with a 50\% gain over an analytic model~\cite{investigationchargesharing}. The distinctive result here is not the core number but the behavior outside the cell. Over the full fiducial the transformer holds a per-axis half-width of \SI{41.1}{\micro m}, while matrix inversion and the $\eta$ formula are at \SI{114}{\micro m} and \SI{168}{\micro m} respectively. The same comparison holds over a wider region. Extended to every matched event out to $|x|,|y| < \SI{0.585}{mm}$, a region $2.3$ pitches across, the transformer still gives a median per-axis residual of \SI{27.6}{\micro m}, or $5.5\%$ of the pitch, and a half-width of \SI{47.2}{\micro m}, or $9.4\%$. As in Ref.~\cite{aoki2026}, these comparisons are not one to one, since the quoted values depend on sensor geometry, readout capacitance, channel count, beam conditions and the definition of the resolution itself.

\section{Conclusions}
\label{sec:conclusion}

We have measured full-waveform machine-learning reconstruction of an AC-LGAD in a minimum-ionizing beam, with the position labels supplied by a telescope aligned and calibrated in situ with \textsc{T\textsuperscript{3}} to about \SI{2.5}{\micro m} per axis so that the residual measures the sensor. The device is a BNL AC-LGAD with a \SI{500}{\micro m} pad pitch and a $2\times2$ readout, exposed to the \SI{120}{GeV} pion beam of the CERN SPS. The topology-agnostic transformer of Ref.~\cite{aoki2026} reconstructs the impact position with a median total residual of \SI{45.9}{\micro m} over a $1.0\times\SI{1.0}{mm\squared}$ fiducial, a factor $2.9$ better than the best analytic estimator on identical events. The advantage narrows to a factor $1.2$ inside the \SI{500}{\micro m} readout cell itself.

The analytic and learned methods are complementary in different regions of the sensor. Near the center, where the response is close to linear, a closed-form charge-imbalance estimator with no free parameters already approaches the observed resolution floor. The network provides no measurable improvement there and has the larger tail. Outside the readout cell, where the charge asymmetry turns over, the learned estimator is decisively better because it need not assume a functional form and can use the shape of the waveform as well as its height. A practical reconstruction can therefore use the analytic estimate as a robust, calibration-light default inside the cell and reserve the network for the region where it improves performance.

The resolution reported in the core is consistent with being limited by the signal-to-noise ratio rather than by the reconstruction, and Eq.~(\ref{eq:snr}) parametrizes that trend; both of its constants are fitted to the eight binned points, so it describes the trend rather than predicting it. At the signal-to-noise ratio achieved here the measurement sits on the steep part of the curve. Higher gain or lower-noise readout therefore provides a measurable resolution improvement, while a larger position-only network does not. The results are also more so limited by the size of the dataset rather than the reconstruction method. Future data-taking with more statistics, more telescopes, higher bias, and finer pitch should yield improved resolution. Timing is also the natural next step. The waveforms already carry timing information, and the architecture is designed to accept it, so extending these results from $(x,y)$ to $(x,y,t)$ can be pursued by adding the target without redesigning the model. The bias-conditioned architecture and topology-agnostic attention provide a path toward deploying one model across operating points and sensor geometries.

\section*{Acknowledgements}
We would like to thank the Solid State Detectors Lab of the EP-DT group at CERN for the support. This work was performed in part within the framework of the DRD3 Collaboration and was supported in part by the Brown University Undergraduate Teaching and Research Award (UTRA), Brown Research Seed Award No. GR300352, and the BEACON 2026 Brown Data Science Collaborative Award. This material is based upon work supported by the U.S. Department of Energy under grants DE-447 SC0012704, DE-SC443363, DE-SC426496 and DE-448 SC0020255. This research used resources of the Center for Functional Nanomaterials, which is a U.S. DOE Office of Science Facility, at Brookhaven National Laboratory under Contract No. DE-SC0012704.

\bibliographystyle{JHEP}
\bibliography{references}

\end{document}